\documentclass[%
 reprint,
 amsmath,amssymb,
 aps, prl
]{revtex4-1}

\usepackage{graphicx}
\usepackage{dcolumn}
\usepackage{bm}
\usepackage{subfigure}

\usepackage{epstopdf}

\begin{document}

\preprint{APS/123-QED}

\title{
         High temperature Bose-Einstein condensation into an excited state at equilibrium}

\author{Raina J. Olsen}
\email{olsenrj@ornl.gov }
\affiliation{Materials Science and Technology Division, Oak Ridge National Laboratory, Oak Ridge, TN 37831, USA}

\begin{abstract}
We describe a new mechanism for high $T_c$ Bose-Einstein condensation. Strongly interacting particles condense into a quantum state which is an excited state at low density, but becomes the ground state as density increases because it minimizes the interaction energy. This strong energetic preference for the condensate is able to overcome the entropic cost of multiple occupation at higher temperatures than traditional systems. Mean field calculations for a graphene potential which holds two closely interacting layers of molecular hydrogen show condensation at temperatures up to 60 K.
\begin{description} 
\item[PACS numbers]
05.30.-d, 67.85.Hj, 03.75.Kk, 67.63.Cd
\end{description}
\end{abstract}

\maketitle

A Bose-Einstein condensate (BEC) is a delicate state of ultra-cold matter in which many boson particles occupy the lowest energy quantum state. While condensation was long suspected to play a role in superfluidity \cite{london}, existence of a BEC was experimentally demonstrated \cite{becRub,becNa,becLi} relatively recently.  Because a BEC only forms when the density of available quantum states (DOS) is reduced to the order of the actual particle density, which typically occurs at temperatures $T\leq$ 2.2 Kelvin (K), sophisticated cooling techniques are required. 

In this letter, we describe a new breed of high temperature BEC which can be observed at more reasonable temperatures.  This high $T_c$ BEC forms from strongly interacting bosons that condense into a quantum state with non-zero kinetic energy, that also minimizes interactions relative to other states, creating a stronger and stronger energetic preference for the condensate as density increases which is able to overcome the entropic cost of multiple occupation at higher temperatures.

\begin{figure}[htb!]
  \centering
\includegraphics[width=0.49\textwidth]{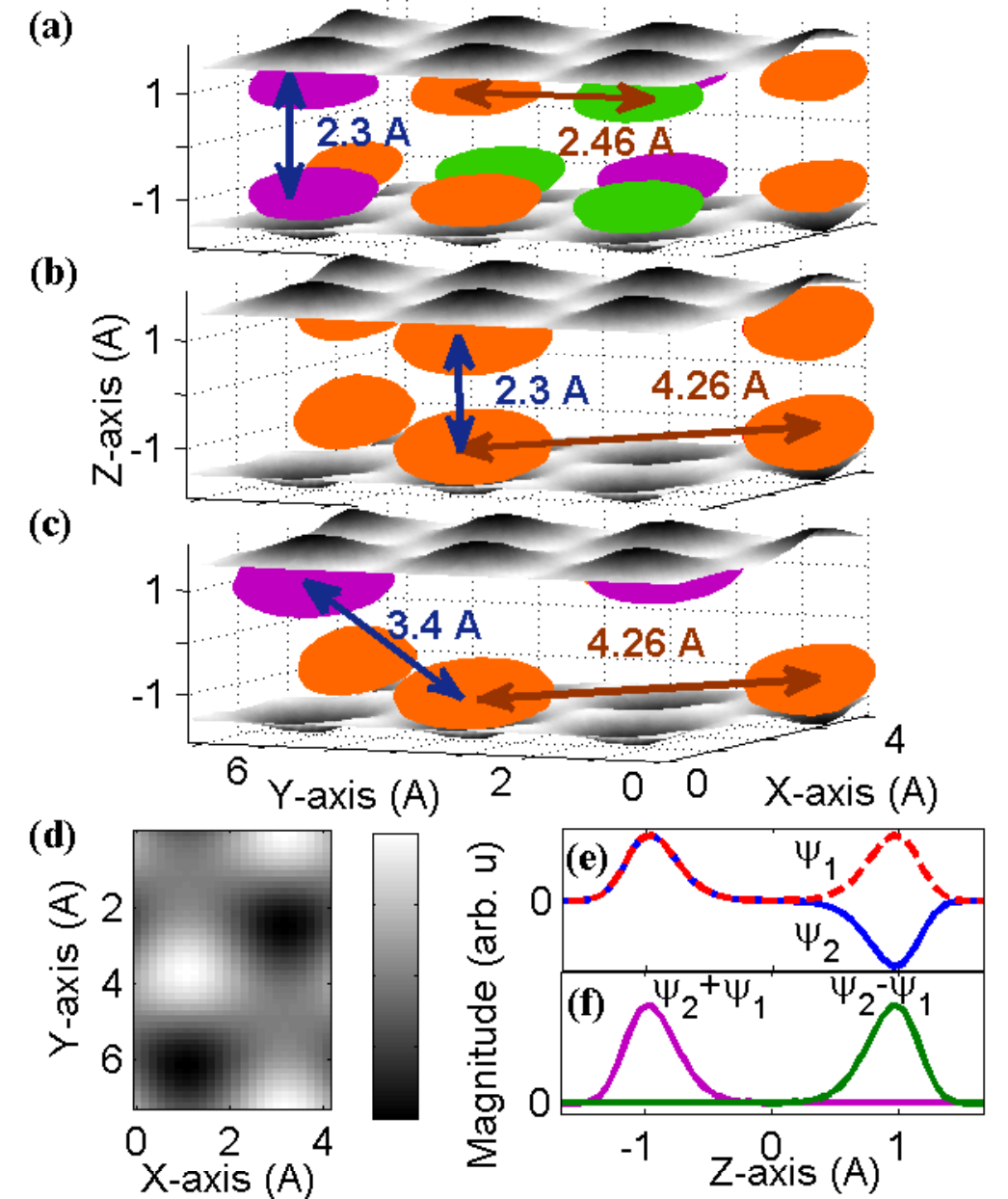}
\caption{Quantum states of H$_2$ in a 8.1 \AA \ wide  graphene slit-pore.   The graphene V=0 potential energy surfaces are shown.  Shaded ovals show regions of concentrated probability density for three H$_2$ states: (a) thermally averaged state   (Color differences aid in distinguishing peaks.) (b) aligned commensurate (AC) state (c) staggered commensurate (SC) state.   Remaining panels depict formation of SC state in Eq. \ref{sc}.  (d)  A wavefunction at the Dirac point.  The magnitude at each graphene hexagon center is alternately +1, 0, and -1.  Hopping between the top and bottom occurs by combination of (e) the first two vibrational wavefunctions.   (f) Adding/subtracting them concentrates the state at the top/bottom. }
\label{potwf}
\end{figure}

While it is conceivable that such high $T_c$ BECs may be observed in many types of systems, we demonstrate the concept here for molecular hydrogen (H$_2$) in a graphene potential with a precisely tuned width, in which H$_2$ forms two closely interacting layers.  For an excited state in which a molecule hops between layers (Fig \ref{potwf}(c)), mean field calculations show the first peak in the pair-correlation function \cite{paircor} for two condensed H$_2$ is increased just past the hard core repulsion diameter \cite{wang}, reducing their interaction enough to produce condensation up to 60 K.  To our knowledge this is the first system for which bosons are predicted to condense into an excited state at equilibrium without being driven by external conditions, as occurs with lasers, condensates of quasi-particle excitations \cite{polariton,magnon}, and vortices \cite{vortex2comp,vortexstir,vortexRot}, solitons \cite{solitonGen,solitonDark}, and phonon-like collective oscillations \cite{collective} evoked in typical BECs. 

\begin{figure*}[htb!]
  \centering
\includegraphics[width=0.9\textwidth]{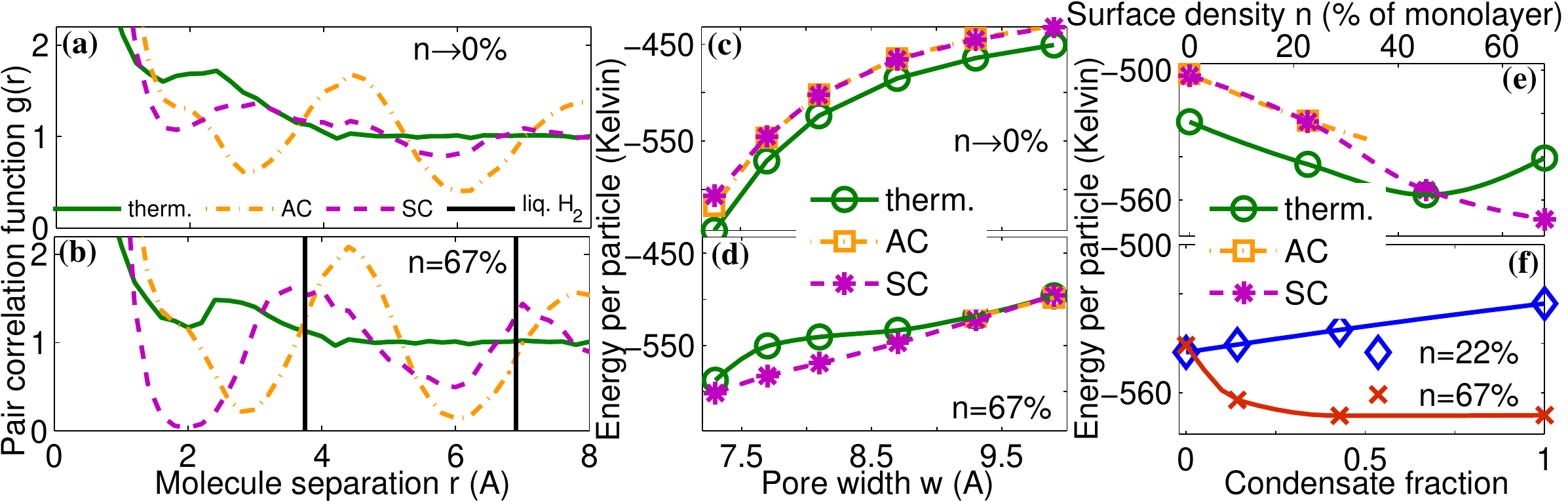}
\caption{Properties of H$_2$ in a thermal distribution of states, or condensed in the aligned (AC) or staggered (SC) commensurate states.  Pair correlation functions for two H$_2$, at densities of (a) $n\rightarrow$0\% and (b) $n=$67\% of a full monolayer of adsorbed H$_2$ for a $w=$8.1 \AA \ pore.  The location of peaks in the pair-correlation function of bulk liquid H$_2$\cite{paircor} are also marked.   Energy per particle as a function of pore width at (c) $n=$0\% and (d) $n=$67 \%.  For the thermal distribution, the ground state energy is given.  Energy per particle for a $w=$8.1 \AA \ pore as a function of (e) density and (f) condensate fraction in the staggered state. }
\label{energies}
\end{figure*}

H$_2$ adsorption potentials created by parallel graphene layers \cite{mattera} with AA stacking separated by width  $w$ are highly corrugated in the potential energy along the plane, by $\sim$45 K.   As a result, the low-energy single-particle quantum states are heavily perturbed from their free equivalents.   These states have peaks in their probability densities which form a triangular lattice with spacing $a=2.46$ \AA.  In pores with  $w\ge$7 \AA, the probability density also separates into two layers, as depicted in Figure \ref{potwf}(a). 

In considering likely forms for many-body quantum states, we rely on neutron scattering experiments \cite{isotopeE, nielsen,nielsenp} showing liquid H$_2$ on graphene at $T=$ 14-34 K tends to have short-range order commensurate with the underlying lattice, meaning the strong influence of the corrugation persists at high density.  The graphene lattice spacing $a$ is smaller than the H$_2$ hard sphere diameter, $d=2.95$ \AA; below this separation the H$_2$-H$_2$ interaction becomes highly repulsive.  Thus only every third lattice point tends to be occupied, with an in-plane nearest neighbor distance of $a'=$4.26 \AA.  The reciprocal lattice vectors of this overlying H$_2$ lattice are equal to the wave-vector at the $K$ and $K'$ points of the graphene lattice.  For $w\simeq$7-10 \AA, the interaction between the top and bottom layers is also significant, and we distinguish between aligned (Fig. \ref{potwf}(b)) and staggered (Fig. \ref{potwf}(c)) lattices.

Matching quantum states may be easily formed from a few single particle states.  We used the \citet{mattera} C-H$_2$ interaction to find the potential $V_{ext}(\vec{x})$ for a graphene pore, and solved the Schrodinger equation numerically \cite{chin} using periodic boundary conditions for surface area $S=32$\AA$^2$.  The resultant quantum states are $\psi_{\vec{k},m,n}$, where $\vec{k}$ gives the wavevector in the first Brillouin Zone, $m$=1, 2$\dots$ is the band number, and $n$=0, 1$\dots$ is the quantum number of the confined motion perpendicular to the plane.  Consider the following probability densities, 
\begin{eqnarray}
\label{thermal} \rho_t&=& \frac{1}{\sum e^{-\beta E_{\vec{k},m,n}}} \sum e^{-\beta E_{\vec{k},m,n}} |\psi_{\vec{k},m,n}|^2, \\
\label{ac} \rho_a&=&|\frac{1}{2}(\psi_{K,1,0} +\psi_{K',1,0})+\frac{1}{\sqrt{2}}\psi_{0,1,0}|^2, \\
\rho_s&=&| \frac{1}{2i}(\psi_{K,1,0} -\psi_{K',1,0})+\frac{1}{\sqrt{2}}\psi_{0,1,1} |^2,
\label{sc}
\end{eqnarray}
where $\beta=1/k_bT$. These are the thermally averaged, aligned commensurate (AC), and staggered commensurate (SC) states depicted in Figure \ref{potwf}(a)-(c).  Figure \ref{potwf}(d) shows the first two terms of Eq. \ref{sc}.  They describe a real sinusoidal wavefunction whose magnitude alternates between +1, 0, and -1 at graphene hexagon centers, which is also in the ground vibrational state.  The last term in Eq. \ref{sc} is the first excited vibrational state.  Because the first two vibrational states (Fig. \ref{potwf}(e))  can be combined to localize the wavefunction at the top or bottom of the pore (Fig. \ref{potwf}(f)), the sum result is to create a state whose probability density hops between the top and bottom.

As Figure \ref{potwf} shows, adjacent peaks in the probability density have the largest separation for the SC state.  In narrow pores where the distance between the two layers $(l\approx w$-5.8 \AA) is small, the SC state is the only one whose adjacent peaks are further apart than $d$ (Fig \ref{energies}(a)).  As a result, the interaction of two H$_2$ in this state is slightly smaller than for any other pair of states.

When a high density of H$_2$ is condensed into the SC state, the coherence between the spatial extent of the H$_2$-H$_2$ interaction and the natural periodicity of the SC state enhances the latter.  

We calculated these many body quantum states using the mean field approximation, neglecting explicit correlation and exchange effects, and finding system energies $E(N,N_a,N_s)$ and states as a function of number of total particles $N$, AC particles $N_a$, and SC particles $N_s$, with the number of particles thermally distributed $N_t=N-N_a-N_s$.  These equations were solved iteratively until self-consistent solutions were found,
\begin{eqnarray}
\label{first} 
\rho^{N,N_a,N_s}(\vec{x})&=&N_t\rho_t(\vec{x}) +N_a \rho_a(\vec{x}) +N_s \rho_s(\vec{x}), \\
\label{dir}
V_{dir}^{N,N_a,N_s}(\vec{x}) &=& \int u[\vec{x},\vec{x}', \rho^{N,N_a,N_s}(\vec{x}')] d\vec{x}' , \\
\hat{H}^{N,N_a,N_s}&=& -\frac{\hbar^2}{2m} \nabla^2 + V_{ext}(\vec{x}) + V_{dir}^{N,N_a,N_s}(\vec{x}),\\
\label{schrod}
E_{\vec{k},m,n}^{N,N_a,N_s} &\psi&_{\vec{k},m,n}(\vec{x})=\hat{H}^{N,N_a,N_s} \psi_{\vec{k},m,n} (\vec{x}), \\
V_{int}^{N,N_a,N_s} &=&\int V_{dir}^{N,N_a,N_s}(\vec{x}) \rho^{N,N_a,N_s}(\vec{x}) d\vec{x}, 
\label{end} 
\end{eqnarray}
\begin{eqnarray}
\label{end} 
E(N,N_a,N_s)&=& N_t E_{0,1,0}^{N,N_a,N_s}+  N_a E_a^{N,N_a,N_s}\\ 
\nonumber + &N_s& E_s^{N,N_a,N_s}-\frac{1}{2} V_{int}^{N,N_a,N_s}. \\
\label{ea}
E_a&=& \frac{1}{2} E_{K,1,0} + \frac{1}{2} E_{0,1,0}\\
\label{es}
E_s&=& \frac{1}{2} E_{K,1,0} + \frac{1}{2} E_{0,1,1}
\end{eqnarray}
Numerical \cite{chin} solutions of Eq. \ref{schrod} are repeated periodically for $S'=15S$ before use in Eqs. \ref{first}-\ref{dir}.  Interactions are included through $u$, a finite-width H$_2$-H$_2$ interaction functional, 
\begin{eqnarray}
 u[\vec{x}, \vec{x}',\rho(\vec{x}')] &=& \\
\nonumber \rho(&\vec{x}')& \times \left \{ \Phi_{LJ}(|\vec{x}-\vec{x}'|), \ |\vec{x}-\vec{x}'| \ge h \atop \Phi_{LJ}(h), \ \ \ \ \ \ \ \ \ |\vec{x}-\vec{x}'| < h \right . \\
\nonumber + \ c \delta &(\vec{x}'&-\vec{x}) \left [  \frac{3}{4 \pi h^3} \int_{|\vec{x}-\vec{r}|\le h} \rho(\vec{x}+\vec{r}) d\vec{r} \right ]^{1+\gamma}, \\
\Phi_{LJ}(r)&=& 4 \epsilon \left [ \left (\frac{\sigma}{r} \right )^{12} -\left (\frac{\sigma}{r} \right )^{6}  \right ],
\end{eqnarray}
similar to those used for liquid $^4$He \cite{finiteFunc,microscopic}.   The second term is increasingly repulsive with density, approximating correlation effects. Table \ref{param} gives parameters:  $\epsilon$ and $\sigma$ are Lennard-Jones parameters \cite{wang}, and the remainder reproduce properties of bulk liquid H$_2$ \cite{deltaFunc,microscopic}.  Due to the relatively low temperatures, we use para-H$_2$ in the spherically symmetric ground rotational state.  While ortho-H$_2$ is still a boson, excited rotational states may require inclusion at higher $T$ in future calculations.

\begin{table}[tb]
\caption[Peak locations.]{
\label{param}
Parameters of the H$_2$-H$_2$ interaction functional.}
\begin{center}
  \begin{tabular}{|c|c|c|c|c|c|}
\hline
parameter & $h$ & $\epsilon$ & $\sigma$ &  $c$& $\gamma$\\
\hline
units  &\AA& K & \AA & K\AA$^{3(1+\gamma)}$ &- \\ 
value &2.93&34.2&2.96&2.19$\times 10^{10}$&4.25\\
\hline
\end{tabular}
\end{center}
\end{table}

We define the surface number density as $n=10.7 N/(A \times 2)$, where 10.7 \AA$^2$ is the surface area per H$_2$ at monolayer coverage \cite{nielsen} and the factor 2 accounts for both layers.  Maximum densities studied were the ones which hold exactly one H$_2$ in each peak of the probability density of the SC and AC states, as long as all peaks are at least $d$ apart.  Above this density, it is certain that H$_2$  are often closer than $d$ and we expect significant correlation effects.  The maximum $n$ at any width was  $n=67$\%, equivalent to one commensurate lattice on each side of the pore.

The pair correlation function for two particles in the SC condensate at this maximum density is shown in Figure \ref{energies}(b).  It is quite similar to that of bulk liquid H$_2$ \cite{paircor}.  (The major exception is the behavior at $r<2$ \AA; this difference originates from the mean field approximation.)  As a result, the interaction of two H$_2$ in this state is much smaller than for any other pair of states.   

Even though the single-particle SC state is always higher in energy than the single-particle ground state (Fig. \ref{energies}(c)), the tendency of the SC state to reduce the interaction energy means that a condensate in this state becomes the many-body ground state as density increases (Fig. \ref{energies}(d)-(f)) for all widths studied.  But it is only within a narrower range, centered around $w\approx$ 8 \AA, that the energy difference between the SC state and a thermal distribution is large enough (Fig. \ref{energies}(d)) to form a condensate at temperatures above the melting point of H$_2$.    In this range, the intermolecular interaction and the SC state are closely enough in phase to reduce the interaction energy significantly.

$N_t$ gives the number of particles that occupy the typical quantum states shown in Fig. \ref{potwf}(a).  The number of these states which are thermally accessible and thus likely to be occupied at $T$ is given by
\begin{eqnarray}
\label{dos}
g_t^{N,N_a,N_s}(T)&=&15 \sum_{\vec{k},m,n} e^{-\beta (E_{\vec{k},m,n}^{N,N_a,N_s}-E_{0,1,0}^{N,N_a,N_s})}\\
&\approx&\frac{S'}{\Lambda^2}=15\frac{S}{\Lambda^2},
\end{eqnarray}
where $\Lambda$ is the thermal de Broglie wavelength.

The canonical partition function was computed for a condensate in the SC state,
\begin{equation}
\label{par}
Z(N,S',\beta)=\sum_{N_s} e^{-\beta E(N,N_a=0,N_s)}  \frac{(N_t+g_t-1)!}{N_t! (g_t-1)!},
\end{equation}
where we have  interpolated between calculated values of $E$ and $g_t$ as as function of $N,N_s$. Bose-Einstein statistics (BES) are incorporated by properly counting the number of microstates for each macrostate $E(N,N_a,N_s)$ \cite{statm2}.  The last term in Eq. \ref{par} gives the number of ways to arrange $N_t$ indistinguishable bosons over $g_t$ states.  Similarly, the number of ways to arrange $N_s$ indistinguishable bosons in the SC state is equal to 1.  Note that while this method correctly uses BES to compute the balance between $N_s$ and $N_t$,  Eqs. \ref{thermal} \& \ref{dos} use Maxwell-Boltzmann statistics to compute the distribution of $N_t$ particles over the $g_t$ states.  This is justified by the fact that for every set of parameters studied, $N_t<g_t$.  A similar condition, $N<V/\Lambda^3$, justifies use of the classical canonical ensemble for a free three dimensional gas \cite{hill}.

The condensate fraction ($f=<N_s>/N$), average system energy, and specific heat were computed from the partition function $Z$.  The temperature (through $\beta=1/k_bT$) was fixed at $T=$ 20 K in Eq. \ref{thermal} and allowed to vary in Eq. \ref{dos}.  (As a check, we performed calculations with $\beta$ variable in Eq. \ref{thermal} for $w=8.1$ \AA, $n=$ 67\%, finding over $T=$ 20-80 K, only a $\leq$8\% variation in the relevant value $E(N_t=N)-E(N_s=N)$.)

\begin{figure}[htb!]
  \centering
\includegraphics[width=0.49\textwidth]{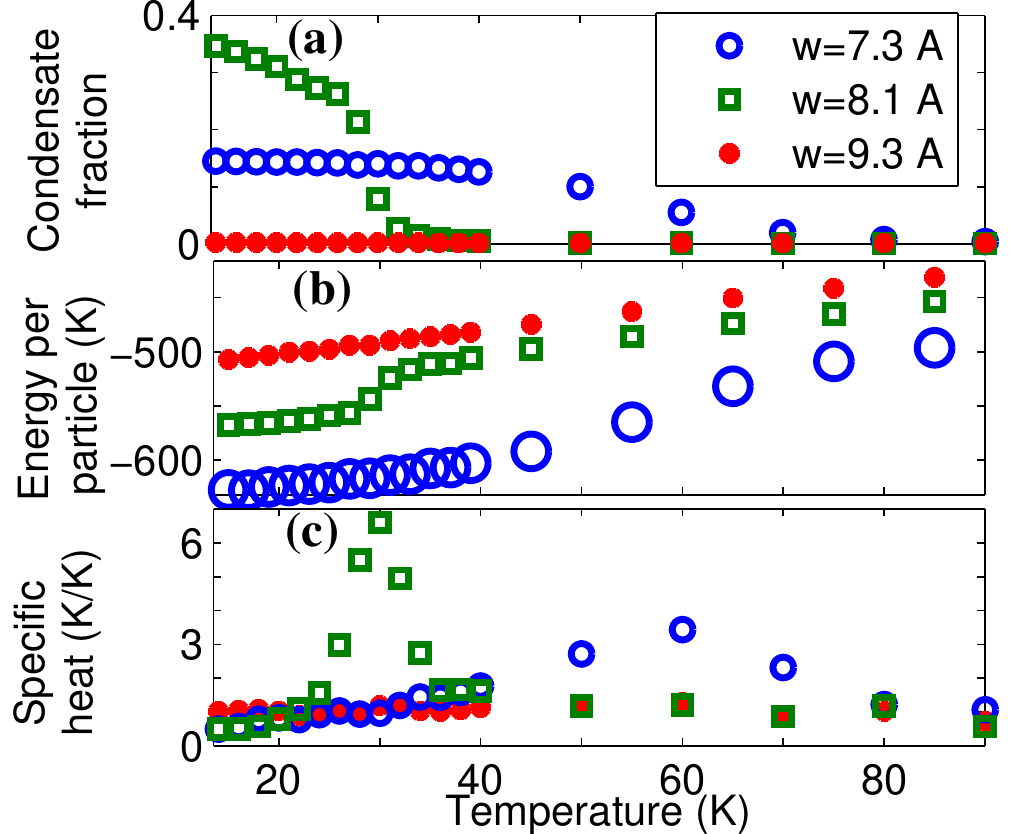}
\caption{(a) Condensate fraction, (b) average energy per particle, and (c) specific heat per particle as a function of temperature at density $n=67$\% for several different pore widths. }
\label{cond}
\end{figure}  

Figure \ref{cond} shows the condensate fraction $f$, average energy per particle, and specific heat as a function of $T$ for several pore widths at a density of $n=67$ \%. Unlike a typical BEC, there is no simple definition for the critical temperature $T_c(w)$.  The condensate fraction has a strong dependence on $n$ (not shown), with condensation only occurring at high density, $n>50$ \%, where the interaction energy is significant.  Nevertheless, the specific heat (Fig. \ref{cond}(c)) has the same characteristic cusp as a typical BEC, and falls toward the constant of a classical two-dimensional gas at $T>T_c(w)$.    

We also found that the $f$ never reaches 1.  This is because the energy is roughly constant above a critical value of $f$, as shown in Fig. \ref{energies}(f).  Above this point, the mean field of the condensate pulls all other states into similar forms which also minimize interactions.  However, we have not included exchange in calculations, a factor which tends to favor condensation in a single state \cite{concepts}.

For the larger pores studied, $w\ge$9.3 \AA, the AC state is lowest in energy at high density, but the energy difference is not enough to form a condensate.   Because the properties of larger pores approach those of independent graphene sheets, this result is consistent with studies of liquid H$_2$ on graphene which have found no evidence of condensation or superfluidity \cite{quantF} except below the melting temperature with impurities \cite{fluidImp}. 

At first glance,  several objections could be made to our claim of high $T_c$ Bose-Einstein condensation in this quasi-two dimensional (2D) system.  Firstly, we claim to see a BEC at temperatures up to 60 K.  But at this temperature, the thermal de Broglie wavelength \cite{hill} or "quantum size" of a hydrogen molecule is $\Lambda=1.5$ \AA, much smaller than the average separation between the particles in liquid H$_2$, 3.7 \AA.  Thus the wavefunctions of adjacent particles should not overlap, and a BEC should not form. However, the de Broglie wavelength is not an intrinsic property of a particle that is independent of its environment, but is derived \cite{hill} from the DOS.  Because of the large gap in energy between the SC state and the other states in this system, there are fewer thermally accessible states, reducing the DOS and increasing $\Lambda$ so that a BEC may form at higher temperatures.

Secondly, it is generally accepted that BECs do not form in 2D.  This was rigorously proven \cite{hohenberg} by considering the phonon spectrum in a 2D system with continuous translational symmetry.  As the phonon wavelengths approach infinity, their energies approach zero, leading to an infrared divergence in the phonon spectrum in 2D which tends to break the condensate apart.  Physically, the long wavelength phonons tend to move macroscopically large sections of the condensate together.   However, the system discussed here has \emph{discrete} translational symmetry, and moving the condensate moves the H$_2$ out of the minima in the graphene potential.  Thus the energy of long-wavelength phonons approaches a finite constant, and there is no infrared divergence.

We also note that because the $H_2$ condense into a single quantum state (SC state), the system has off-diagonal long range order, with the \emph{average} value of the first-order coherence function $g(r)$ approaching a finite value as $r\rightarrow \infty$. But because of the natural periodicity of the SC state, $g(r)$ is not constant, but oscillates.

Here we have described a new mechanism for creating a Bose-Einstein condensate whose critical temperature is an order of magnitude higher than other BECs formed from real particles.  Development of high $T_c$ BECs would reduce the sophistication and cost of the cooling techniques required, and also permit higher densities by eliminating the need of super-cooling.  Because the confining potential described here is solid state (thus static) and evaporative cooling is not required, there should also be no limitation on the lifetime of the condensate.

While we have presented calculations for a high $T_c$ BEC of H$_2$  in graphene pores,  other species and/or potentials may be substituted, as long as the same basic concept is employed:  the periodicity of the potential and the inter-particle interactions must be tuned together so that a small subset of quantum states minimize the interactions.  In particular, if this new BEC can be created from molecules in an optical lattice \cite{oplatreview} with tunable interactions \cite{reviewtuneopl}, the resultant system would have the advantages of a high $T_c$ BEC, in addition to the controllability and lack of defects of an optical lattice potential.  An optical version could also be studied with simple light scattering rather than neutron scattering \cite{dinsHe,dinsHeAds}, as required for the opaque H$_2$-graphene system.

Even with the limitaions of real materials, the H$_2$-graphene system may provide immediate benefits for hydrogen storage \cite{H2storage} and sorption cryo-coolers \cite{sorptionCool}.  A high $T_c$ optical version might someday be useful for large scale, cost-effective applications like quantum computing which cannot feasibly operate at ultra-cold temperatures.

\section*{Acknowledgments}
This work was supported by the U.S. Department of Energy Energy Efficiency and Renewable Energy (DOE-EERE) Postdoctoral Research Awards under the EERE Fuel Cell Technologies Program, administered by ORISE for the DOE. ORISE is managed by ORAU (DEAC05-06OR23100).  We would like to thank J. Morris, V. Cooper, and G. Vignale for helpful discussions.

\end{document}